# Cellular Uptake and Biocompatibility of Bismuth Ferrite Harmonic Advanced Nanoparticles

Davide Staedler[1, a]; Solène Passemard[1, a]; Thibaud Magouroux[b]; Andrii Rogov[b]; Ciaran Manus Maguire[c]; Bashir M. Mohamed[c]; Sebastian Schwung[d]; Daniel Rytz[d]; Thomas Jüstel[e]; Stéphanie Hwu[b]; Yannick Mugnier[f]; Ronan Le Dantec[f]; Yuri Volkov[c,g]; Sandrine Gerber-Lemaire[a]; Adriele Prina-Mello[c,g]; Luigi Bonacina[b,*] and Jean-Pierre Wolf[b]

[1]These authors contributed equally to this work.

[a]Institute of Chemical Sciences and Engineering, EPFL, Batochime, 1015, Lausanne, Switzerland.

[b]GAP-Biophotonics, Université de Genève, 22 Chemin de Pinchat, 1211 Genève 4, Switzerland.

[c]Nanomedicine Laboratory and Molecular Imaging group, School of Medicine, Trinity Centre for Health Sciences, Trinity College, D8, Dublin, Ireland.

[d]FEE Gmbh, Struthstrasse 2, 55743 Idar-Oberstein, Germany.

[e]Fachbereich Chemieingenieurwesen, Fachhochschule Münster, Stegerwaldstrasse 39, 48565 Steinfurt, Germany.

[f]Univ. Savoie, SYMME, F-74000 Annecy, France.

[g]CRANN Naughton Institute, Trinity College, D2, Dublin, Ireland.

*Corresponding author*

**Funded by:** Partially funded by European Commission funded project NAMDIATREAM project (FP7 LSP ref 246479) and CAN project (European Regional Development Fund through the Ireland Wales Programme 2007-13 INTERREG 4A);

**Conflicts of interest:** the authors declare no conflict of interest.

## Abstract

Bismuth Ferrite (BFO) nanoparticles (BFO-NP) display interesting optical (nonlinear response) and magnetic properties which make them amenable for bio-oriented applications as intra- and extra membrane contrast agents. Due to the relatively recent availability of this material in well dispersed nanometric form, its biocompatibility was not known to date. In this study, we present a thorough assessment of the effects of *in vitro* exposure of human adenocarcinoma (A549), lung squamous carcinoma (NCI-H520), and acute monocytic leukemia (THP-1) cell lines to uncoated and poly(ethylene glycol)-coated BFO-NP in the form of cytotoxicity, haemolytic response and biocompatibility. Our results support the attractiveness of the functional-BFO towards biomedical applications focused on advanced diagnostic imaging.

**Background**

Most of nanophotonics approaches (quantum dots, plasmonic nanoparticles (NP), up-conversion NP) for health applications present static optical properties (absorption bands, surface plasmon resonances) often in the UV-visible spectral region and do not fully allow for exploiting the tuning capabilities of new laser sources and their latest extensions in the infrared. To circumvent these limitations, a few research groups in the last years have introduced a new nanotechnology approach based on inorganic nanocrystals with non-centrosymmetric structures. Such nanomaterials present a very efficient nonlinear response, and can be easily imaged by their second harmonic generation (SHG) in multi-photon imaging platforms. [1-6] Such harmonic NP (HNP) do not suffer from conventional optical limitations such as photobleaching and blinking allowing long-term monitoring of developing tissues [4, 7]. Several HNP have been recently synthesized and tested for biological applications [3-5, 7]. Particular care should be paid to assess the ability of these NP to reach intracellular targets without causing major interferences to the cell metabolism. In this context, this subcellular targeting becomes increasingly important as key parameters for the understating of complex events in living cells [8, 9]. In fact, the possibility to freely change detection wavelength can be exploited for subtle co-localization studies with organelle-specific dyes, as the signal from NP can always be selectively told adapt. However, one aspect that is not fully understood and remains uncertain is how nanomaterials interact with cellular interfaces such as cytoskeletal membranes since it is known that small alterations in their physicochemical properties can drastically influence the cells-NP interactions, especially the uptake mechanisms [9, 10]. Therefore, lead-NP-candidate identification process based on high throughput screening as decision-making process is a prerequisite for the validation of new SHG NP for bio-imaging applications. Here we present a study based on $BiFeO_3$ (bismuth ferrite, abbreviated as BFO) NP (BFO-NP), which were recently successfully introduced as photodynamic tools and imaging probes [11]. Nonetheless, such is the technological novelty of this new group of materials that there is still a knowledge gap that requires the scientific community attention towards the investigation of the interaction at the cellular and subcellular levels. The opportunity of closing this gap is presented by providing the first thorough investigation on the effects of BFO-NP in cellular metabolism and uptake mechanisms. Toxicity and biocompatibility were assessed by automated high content screening, recording cytotoxicity, lysosomal mass and cell permeability, in line with previously published works [12, 13]. Cellular uptake was investigated by co-localizing the NP with specific fluorophores for cell membranes and endosomes. Moreover, in this paper we present for the first time to our knowledge the most efficient protocol for the coating of these HNP with poly(ethylene glycol) (PEG) derivatives to promote colloidal stability and biocompatibility in biological media, and to allow post-functionalization with bioactive molecules[14, 15]. In this context, the biocompatibility, cellular uptake and intracellular localization of free and PEG coated BFO-NP were compared.

**Methods**

*Preparation of a polydisperse suspension of BFO*

The starting BFO suspension (lot BFO0018, 62.5 wt %) in $ZrO_2$ balls was provided by the company FEE (Germany) under a collaboration agreement. Two mL of the suspension were taken off and diluted in 2 L of EtOH. The suspended nanoparticles were ultra-sonicated overnight (O/N) and sedimented for 10 days. Fifty mL of the upper polydisperse suspension were transferred in a round bottom flask, 4 mL of oleic acid was added and EtOH was removed under vacuum. The residue was weighed and suspended in EtOH to obtain a stock solution at 3.6 mg.mL$^{-1}$.

*Coating of BFO-NP*

BFO-NP (solution in EtOH, 3.6 mg/mL, 583 µL) were diluted in 1 mL of a mixture of EtOH:toluene 1:1 and aqueous ammonia (25%, 320 µL) was added. The suspension was ultra-sonicated for 30 min. Amino silane-PEG (**2**) and azido silane-PEG (**1**), synthetized as described in ref.13 (ratio 1:1, 43 µmol, 100 mg) were added and the suspension was ultra-sonicated at 40 °C for 16 hr. The suspension was then concentrated under vacuum to a small volume and distributed in plastic tubes. To each Eppendorf was added a mixture of dichloromethane (DCM):EtOH:water (1:1:1, 1 mL) and the solutions were shaken until emulsion. The emulsion was broken by centrifugation (10 min, 13 000 rpm) and a DCM layer showed the presence of a slightly orange suspension, which corresponded to coated BFO-NP. The aqueous layer, containing the excess of unreacted polymers, was removed. To each Eppendorf was added a mixture EtOH:water (1:1, 0.5 mL) and the solutions were shaken until emulsion, then centrifugated (10 min, 13 000 rpm). The procedure was repeated 5 times to obtain pure suspension of BFO-NP in DCM. The organic solvent can be easily removed in vacuum and replaced by EtOH. The BFO-NP concentration was calculated by measuring the turbidity of the solutions by spectrometry at 600 nm (Synergy HT) and by comparing the values with a standard-curve prepared using the stock solution at 3.6 mg/mL.

*Characterization of uncoated and coated BFO-NP*

Advanced physico-chemical characterization of BFO-NP was recently performed[16]. In this work BFO-NP were characterized using a Zetasizer NanoZ (Malvern) for the measurement of the dynamic light scattering (DLS) and the zeta potential. Suspensions of uncoated or coated BFO-NP (20 µL) were diluted in 1 mL of distilled water. Acetic acid (100 µL) was added and the resulting suspensions were ultra-sonicated for 30 min and analysed.

*Nanoparticles characterization in biological media*

The physico-chemical characterization of the NP was carried out by nanoparticle tracking analysis (NTA). BFO-NP at 25 µg/ml were vortexed for 5s to disperse the particles and then diluted at 1 µg/mL in different solutions (0.22 µm filtered): diethylpyrocarbonate (DEPC) water, Dulbecco's Modified Eagle Medium (DMEM), Ham's F-12K (Kaighn's) Medium (F12K) and Roswell Park Memorial Institute (RPMI) culture media, and their supplemented form with 10% fetal bovine serum (FBS). The dispersions were then analyzed via NTA for the physico-chemical characterization measurement of hydrodynamic radius and polydispersity index (PDI) at room temperature (RT) of individual BFO-NP. NTA measurement was done using a Nanosight NS500 (Nanosight). The device consists of an EMCCD camera (ANDOR) mounted on a conventional optical microscope with a 20x objective and LM14 viewing unit containing a 532 nm continuous wave laser light source. Introduction of each sample into the LM14 viewing unit was automated via an on board peristaltic pump with manual focus of nanoparticle following the manufacturer's standard operating procedures. The NanoSight NS500 recorded six independent 90 seconds videos containing fresh nanoparticle populations with each recording. Analysis was conducted in batch mode and analysed with the NTA 2.3 software. All measurements were carried out three times at physiologically relevant pH (pH = 7.4) and means and standard deviations (SD) were calculated. Quality assurance over the measurements carried out was guaranteed by the adoption of Quality Nano (QNano, FP7 project) standard operating procedures (SOPs), which have been developed as part of large inter laboratory comparative study focused on nanoparticle physico-chemical characterization[17].

*Cell model and culturing conditions*

Human lung-derived A549 and NCI-H520 cancer cell lines and human monocytic THP-1 cell line are available from ATCC (American Tissue Culture Collection, Manassas, VA, USA). A549 were grown in DMEM medium containing 4.5 g/L glucose, 10% FBS and penicillin/streptomycin (PS) in a 37 ˚C incubator with 5 % $CO_2$ at 95 % humidity. NCI-H520 and THP-1 were grown in complete Roswell Park Memorial Institute (RPMI) 1640 medium supplemented with 10% FCS and PS in a 37 ˚C incubator with 5 % $CO_2$ at 95 % humidity. For differentiation into macrophages, THP-1 cells were plated at a density of 20'000 cells/cm$^2$ in RPMI 1640 supplemented with 10% FBS and 100 ng/ml phorbol 12-myristate 13-acetate (PMA, Sigma-Aldrich) for 72h. Differentiated THP-1 cells adhered to the bottom of the wells.

*Fluorescent staining for cellular imaging*

The cells were grown for 48h or 72h for activated PMA THP-1 cells, in a 24 well plate containing one rod-shaped microscope slide (BD Falcon). After this time the cells were exposed to BFO-NP at 25 µg/mL or to vehicle (ethanol) at indicated time. For endosomes imaging the cell layers were exposed for 2h to 15 µg/mL of the fluorescent probe FM1-43FX (Invitrogen). After incubation, the cells were fixed in 4% formaldehyde in PBS for 30 min, then washed once in 0.1% Triton X-100 in PBS and twice in PBS, then maintained in formaldehyde. For membrane staining, cells were fixed and then exposed to 0.1 µg/mL Nile Red (Invitrogen) in PBS for 5 min, then rinsed twice in PBS and maintained in formaldehyde. Cell nuclei were stained with 4',6'-diamidino-2-phenylindole (DAPI).

*Multiphoton Laser Scanning Microscopy*

The cells were observed using a Nikon multiphoton inverted microscope (A1R-MP) coupled with a Mai-Tai tunable Ti:sapphire oscillator from Spectra-Physics (100 fs, 80 MHz, 700-1000 nm). A Plan APO 40× WI N.A. 1.25 objective was used to focus the excitation laser and to epi-collect the SHG signal and dye markers fluorescence. Nanoparticles and fluorescent dyes (FM1-43FX and Nile Red) were excited at 790 nm and observed through tailored pairs of dichroic mirrors and interferometric filters (Semrock, FF01- 395/11-25 for SH, FF01-607/70-25 for fluorescence). Statistics were calculated by averaging measured values from samples between 40 and 170 cells per condition for cell labelling and between 8 and 100 cells per condition for co-localization measurements. Cell labelling was analysed by dividing the number of cells labelled by at least one nanoparticle to the total number of cells in a microscopy field and expressed as % of total cells. Co-localization with endosomes was estimated by counting the number of nanoparticles co-localizing with the fluorescence signal from cell membrane dye FM1-43FX using Nikon Imaging Software (NIS) and dividing it by the total number of nanoparticles located inside each cell and expressed as % of all particles internalized in each cell.

*In vitro dose and exposure endpoint determination*

Cytotoxicity on three-cell line models was investigated in vitro after 24 h and 72 h incubation with BFO-NP. Following standardization of the BFO preparation protocols, all were injected into 96 well plates to a final volume of 200 mL/well. A549, NCI-H520 and activated PMA THP-1 cells were incubated with 1.0, 2.5, 5.0, 7.5 or 10 µg/mL of uncoated and coated BFO-NP for 24 h and 72 h in a 37 ˚C incubator with 5 % $CO_2$ at 95 % humidity. Experiments were repeated three times, using triplicate wells each time for each formulation tested. Positive and negative controls were also included into each experiment in order to quantify the extent of toxicity response induced by each particle. Three positive controls were Valinomycin (VAL,

Fisher Scientific) (final concentration 120 μM) to measure changes in mitochondrial transmembrane potential, Tacrine (TAC, Sigma-Aldrich) (final concentration 100 μM) to measure changes in lysosomal mass/pH and Quantum Dots (CdSe) (final concentration at 1 μM) to measure nanoparticle-induced uptake [12]. After 24 h and 72 h incubation, cells were washed in phosphate-buffered saline solution (PBS) at pH 7.4 and fixed in 3 % paraformaldehyde (PFA). A multiparametric cytotoxicity assay was performed using the Cellomics® HCS reagent HitKitTM as per manufacturer's instructions (Thermo Fisher Scientific Inc.). For each experiment, each plate well was scanned and acquired in a stereology configuration of 6 randomly selected fields. In total, each endpoint data plot in the heatmaps represents the analysis of an average of 270,000 cells = 3 runs x 3 triplicates x 6 fields x 5000 cells (on average from t = 0 h). Images were acquired at 10x magnification using three detection channels with different excitation filters. These included a DAPI filter (channel 1), which detected blue fluorescence of the Cellomics® Hoechst 33342 probe indicating nuclear intensity at a wavelength of 461 nm; FITC filter (channel 2), which detected green fluorescence of the Cellomics® cell permeability probe indicating cell permeability at a wavelength of 509 nm and a TRITC filter (channel 3), which detected the lysosomal mass and pH changes of the Cellomics® LysoTracker probe with red fluorescence at a wavelength of 599 nm.

*Statistical analysis*

Response of each cell type to the coated and uncoated BFO-NP was analysed by 2-way ANOVA with Bonferroni post-test analysis. A p-value <0.05 was considered to be statistically significant. In this work we are comparing 4 cell parameters associated with the cytotoxicity response of 3 cell lines exposed to uncoated or coated BFO, at 5 doses, with 3 controls; therefore, the statistical value associated with our work carried out by High Content

Screening and data mining is of significance. To visualize the data, KNIME (http://KNIME.org, 2.0.3) data exploration platform and the screening module HiTS (http://code.google.com/p/hits, 0.3.0) were used. Knime is a modular open-source data manipulation and visualization programme, as previously reported[12, 18-20]. All measured parameters were normalized using the per cent of the positive controls. Z score was used for scoring the normalized values. These scores were summarized using the mean function as follows Z score (x-mean)/SD, as from previous work[12, 18-20]. Heatmaps graphical illustration in a colorimetric gradient table format was adopted as the most suitable schematic representation to report on any statistical significance and variation from normalized controls based on their Z score value. Heatmap tables illustrate the range of variation of each quantified parameter from the minimum (green) through the mean (yellow) to the maximum (red) according to the parameter under analysis.

*Haemolysis assay*

Fresh human blood in lithium heparin-containing tubes was obtained from leftovers of analytical blood with normal values. The plasma was removed by centrifugation for 10 min at 2500 rpm and the blood cells were washed three times with sterile isotonic PBS solution, then diluted 1:10 in PBS. Cell suspensions (300 µL) were added to 1200 µL of each solution containing NP or chemicals in human plasma or PBS at indicated concentrations. Nanopure water (1200 µL) was used as a positive control and human plasma or PBS (1200 µL) were used as negative controls. The mixtures were gently mixed then kept for 2h at RT, centrifuged for 2 min at 4000 rpm and the absorbance of the upper layers was measured at 540 nm in an absorbance multi-well plate reader (Synergy HT). The percentage of haemolysis of the samples was calculated by dividing the difference in absorbance between the samples and the negative control by the difference in absorbance between the positive and negative controls.

Experiments were conducted in triplicate wells and repeated twice. Means ± SD were calculated.

## Results

As previously stated, this work presents for the first time several important aspects relevant to the translation of BFP-NP into a novel biomedical imaging probe for advanced diagnostic screening.

### *Coating and characterization of BFO-NP*

The presence of reactive hydroxyl groups at the surface of BFO-NP facilitates the surface coating chemistry. Modified poly(ethylene glycol) (PEG) 2000 (molecular weight of 2000 g/mol, 45 units) containing silane anchoring groups and reactive functionalities were synthetized as previously published[15] and used to perform covalent coating via silane (Si) ligation (Figure 1).

A previously published study, by some of the authors, focused on the surface coating and post-functionalization of metal oxide NP such as iron oxide NP[15]. In this work, BFO-NP were treated with an equimolar mixture of α- triethoxysilyl-ω-azido and α- triethoxysilyl-ω-amino PEG oligomers **1** and **2**, prepared from linear PEG 2000, in the presence of aqueous ammonia[15]. Ultra-sonication at 40°C for 16 hours, followed by repetitive cycles of decantation/centrifugation into 1:1:1 DCM:EtOH:water resulted in coated BFO-NP (PEG-BFO-NP), which were suspended in EtOH for further characterization. Efficient coating was proved by FT-IR analysis (Supplementary Figure 1). Size and surface charge characteristics of uncoated BFO-NP (U-BFO-NP) and PEG-BFO-NP were measured using DLS and zeta potential techniques as previously described[15]. Upon coating, the zeta potential value shifted from -29.0 ± 1.3 mV to -9.8 ± 0.3 mV and the mean hydrodynamic diameter decreased from 128.8 ± 11.2 nm to 96.1 ± 8.3 nm. The decrease in the hydrodynamic diameter of coated

BFO-NP can be attributed to a better dispersion of the NP in the solvent as a result of a possible colloidal stabilisation [21].

*BFO-NP stability and characterization in biological media*

For both uncoated and coated BFO-NP, extended physico-chemical characterization was carried out after ultracentrifugation and re-dispersion in ultrapure deionized water prior to incubation into relevant biological dispersing media. The two devised BFO-NP potential probes were also characterized by NTA, aiming at the identification of hydrodynamic radii, the colloidal and the aggregation stabilities at physiologically relevant conditions (Figure 2). Interestingly, U-BFO-NP formed aggregates in biological relevant media, particularly when the media were supplemented with serum. PEG-BFO-NP resulted in a decrease of aggregate formation, and thus a better stability of the suspension. This finding was further confirmed after 24h incubation (supplementary Figure 2)

*Interaction between BFO-NP and human derived cells*

The interactions between U-BFO-NP and PEG-BFO-NP were then studied in three human-derived cell lines, one human adenocarcinoma cell line (A549) derived from alveolar epithelial type II cells, one lung squamous carcinoma cell line (NCI-H520) [5] and one human acute monocytic leukemia cell line (THP-1) as a model for macrophages [22]. These experiments were aimed to observe the interaction between particles and cell membranes *in vitro* and to explore the uptake of particles in cells, particularly the endocytic pathways. Indeed, endocytosis represents one of the main internalization mechanisms of NP in cells [10, 14, 23], particularly in macrophages [23, 24]. For the quantification of particles associated with the plasma membranes and internalized into the cytoplasm, cells were exposed to BFO-NP, fixed and labelled with a fluorescent probe used to stain membranes and lipids [25]. HNP uptake by endocytosis was observed by co-localizing the SHG signal with a molecular probe specifically internalized in the endosomes [26]. It is worth pointing out that for the following uptake quantification we purposely adopted an extremely strong criterion (a cell is labelled if at least one particle is in contact with it). Even with this strict definition, we observed significant difference in uptake between 2 h and 24 h, with the notable exception of the macrophages. In fact, a consistent uptake by these latter cells was observed after short exposures (2 h and 24 h), whereas in human lung-derived NCI-H520 cancer cells the uptake of the HNP was only observed after 72 h incubation (Figure 3). HNP internalization in human lung-derived cancer cells is exposure-time dependent (Figure 3). Indeed, HNP generally adhere to the cell membrane after 2 h and are then internalized when increasing exposure time. After 72 h exposure, BFO-NP formed aggregates in intracellular organelles, such as lysosome or endosomes. We did not observe a statistically significant difference in labelling between PEG-BFO-NP and U-BFO-NP after 2 h or 24 h exposure (Figure 3). Interestingly the interaction HNP with immune THP-1 cells was fairly rapid, as more than 80% of cells were found to have internalized at least one particle already after 2 h exposure. Such finding is in line with previous observations, since it is known that THP-1 phagocytic cells rapidly envelop and digest extraneous objects as initial response to infection or foreign body invasion [22, 27, 28]. Conversely, the labelling of A549 and NCI-H520 cells showed a clear exposure dependent behaviour. After 72 h, the number of labelled cells was significantly higher in the presence of PEG-BFO-NP than with U-BFO-NP (Figure 3), suggesting the existence of an exocytosis mechanism for the uncoated BFO-NP as already observed and reported for others metallic NP [29, 30]. In macrophages, the uptake of uncoated particles is mainly endocytic-dependent, whereas the PEGylation of BFO-NP completely reduced the endocytic-mediated uptake (Figure 3). This mechanism of particle internalization was confirmed by multiphoton laser scanning microscopy z-stack analysis of differentiated THP-1, which showed clusters of BFO-NP into the cytoplasm of cells having internalized uncoated BFO-NP after 24 h exposure (Figure 4) as opposed to the behaviour observed with coated BFO-NP, which showed a reduced uptake. In the lung-derived cancer cell lines only a weak fraction of HNP co-localized with the endosomes after 2 h and 24 h exposure, nonetheless this portion increased for the NCI-H520 cells after 72 h exposure to the particles.

*Multiparametric cytotoxicity evaluation of BFO-NP*

The cytotoxicity of uncoated and coated BFO-NP was investigated *in vitro* on the two human lung-derived NCI-H520 and A549 cancer cell lines and the macrophage-derived THP-1 cells, after 24 h and 72 h exposure to increasing concentrations (1, 2.5, 5, 7.5 and 10 μg/ml) of particles. Experiments were repeated three times, using triplicate wells each time for each formulation tested, as shown in supplementary Table 1. Three key parameters of toxicity-attributed phenomena were analysed during this assay: cell number, lysosomal mass and pH, and cell membrane permeability, as previously reported [12]. Indeed, it is known that some toxins can interfere with the cell's functionality by affecting the pH of organelles such as lysosomes and endosomes, or by causing an increase in the number of lysosomes present [12, 31]. Cell membrane permeability changes can be measured as enhancement of cell membrane damage and decreased cell viability as result of cell nuclear-staining counting reduction [12, 32]. Moreover, three positive controls were introduced in this analysis: CdSe quantum dots (QD), which are established toxic NP [12], Valinomycin (Vac), an inducer of energy-dependent mitochondrial swelling causing cell membrane permeability [33] and Tacrine (Tac), a reversible cholinesterase inhibitor which causes increases in cellular lysosome content and lysosomal mass [34]. The rate of cell viability and proliferation was assessed by automated quantitative analysis of the nuclear count and cellular morphology; in parallel to that the fluorescent staining intensities reflecting cell permeability and lysosomal mass/pH changes were also quantified for each individual cell measured within each imaged field and then colorimetrically normalised against the respective controls (Figure 5). A multiparametric analysis of the cellular responses allowed the identification of the different toxicity levels between the two BFO-NP. In general, BFO-NP in concentration range of 1-10 μg/ml showed low cytotoxicity on the cell models adopted. In agreement with the previous results, THP-1 cells are the most affected cells between the three cell lines since these are the first line immune response to any foreign objects such as nanoparticles. For both PEG-BFO-NP and U-BFO-NP a time-dependent and dose-dependent toxicity was measured. However, these effects were more pronounced when cells were exposed to uncoated particles. Among the parameters measured during the NP comparison, the cell permeability parameter showed the most pronounced difference, suggesting that BFO-NP, particularly uncoated particles, interfered mainly with the physiology of the plasma membrane, as also shown by the Vac response as positive control for cellular permeability. PEG-BFO-NP got stored in lysosomal compartments as from lysosomal concentration response, which was also comparable to the response to Tac used positive control for lysosomal response.

*Haemolysis assay*

After having assessed the colloidal stability of PEG-coated BFO-NP in biological media, the interactions with cell membranes were tested by assessing their haemolytic effect on human red blood cells (HRBc) according to established protocols [5, 35]. Haemolysis is defined as the destruction of red blood cells and it is regarded as a key parameter for the evaluation of NP biocompatibility [5, 35, 36]. NP can exert haemolytic effect by electrostatic interactions with membrane proteins or by other NP-specific mechanisms such as the generation of reactive oxygen species (ROS), causing irreversible damage to cells. The assessment of the haemolysis of uncoated BFO-NP showed a weak haemolytic potential (Figure 6), comparable to that of metallic NP observed in other bio-assays [36, 37]. Upon PEG coating, this potential was significantly reduced ($p < 0.001$, Figure 6). These statistically significant results proved that PEGylation of the metal oxide core contributed to the reduction of the interaction between cell membranes and particles surface.

**Discussion**

BFO-NP are a class of nanomaterials with unique optical and physical properties attracting the interest of researchers for several technological applications, in particular related to their multiferroics nature. We have shown recently their potential in the context of advanced non-linear optical imaging[16]. Recently BFO-NP were employed *in vitro* in a pilot study where they proved to be able to locally induce DNA damages by deep UV generation [11]. The cytotoxicity, haemolytic response and internalization mechanisms evidence here reported for coated BFO-NP suggested good biocompatibility and a great potential for biomedical imaging in diagnostic applications. Therefore, the aim of this work was, for the first time, to present the behaviour of these particles in their uncoated or PEGylated form. Such assessment comes timely after the first demonstrations of their interest for bio-imaging and selective photointeraction, as, due to their novelty, there is still a knowledge gap that requires the scientific community attention towards the investigation of their biological effect. NP must be coated with biocompatible polymers in order to stabilize the NP suspensions in biological media and to increase their biocompatibility [38, 39]. Coating with organic polymers also allows particles conjugation and functionalization with biologically-active ligands, such as targeting-specific ligands, therapeutic agents, peptides or antibodies[15, 40]. Here we present a method for the successful PEGylation of BFO-NP based on heterobifunctional PEG oligomers, for which a fast and convenient large scale synthesis protocol was recently published [15]. PEG was selected as coating polymer due to its interesting properties for biomedical applications. Indeed, when compared to any other known polymer, PEG exhibits high hydrophilicity, low protein adsorption, low uptake by immune cells, and no toxicity properties [14, 41, 42]. Our findings confirmed the important role of this kind of coating in terms of biocompatibility. Interestingly as presented here, PEG-BFO-NP are less toxic and their uptake in immune-responsive THP-1 cells is reduced. As a result of the multiparametric cytotoxicity evaluation, we hypothesize that BFO-NP toxicity is mainly mediated by electrostatic interactions with the surface of the particles and the cell membranes. The reduction of cytotoxicity observed upon coating is probably due to a steric barrier between the surface of the coated particles and the lipid membrane of cells [14, 15, 41]. PEG-BFO-NP were internalized in intracellular organelles and they remained located into the cells sensibly longer than U-BFO-NP. This opens interesting ways for biomedical applications, which require stable incorporation of HNP in cells. In the future, more detailed studies about BFO-NP co-localization in cells are needed in order to better understand the pathways involved in the internalization of these particles. Moreover, more detailed assessment of the immune response associated with these particles, such as a cytokine profile after exposure to the particles, could be interesting for the elucidation of their biological effects since they are immune-responding to foreign body [43-45]. However, this work confirmed the biocompatibility of BFO-NP and their utility as nanoprobes for biological applications.

**Figures**

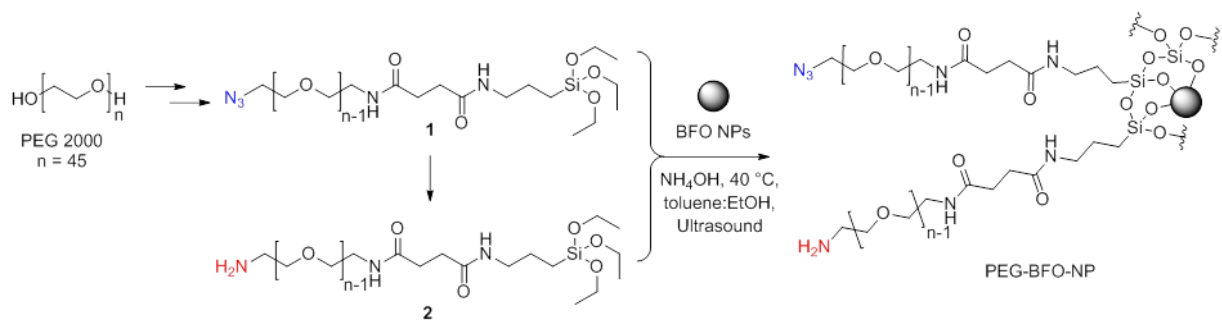

**Figure 1**: Synthesis of coated BFO-NP. PEG oligomer (1) and PEG oligomer (2) were synthetized from PEG 2000 (MW 2000 Da) as published[15], then used for the coating of BFO-NP.

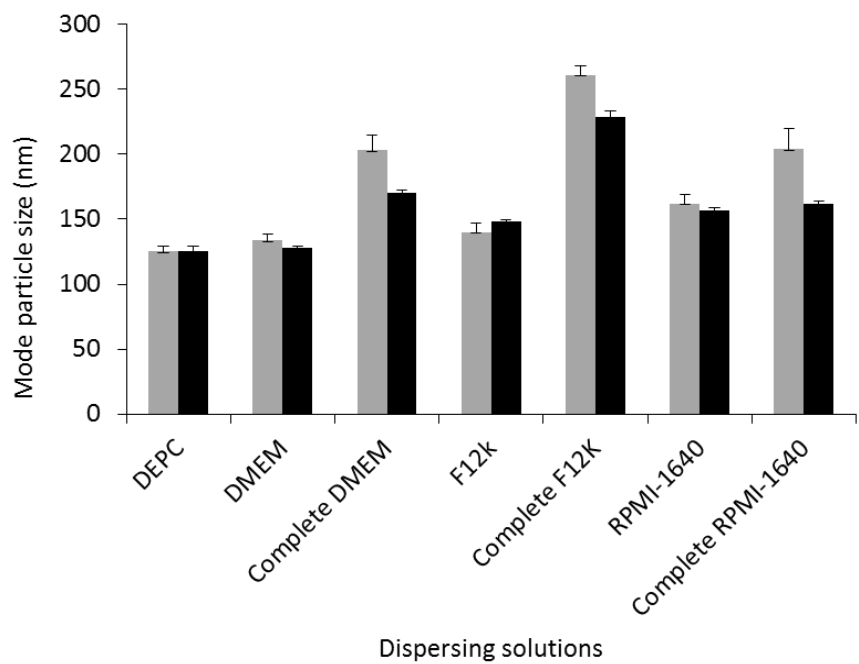

**Figure 2**: Nanoparticle Tracking Analysis of U-BFO-NP (light grey bars) and PEG-BFO-NP (dark grey bars) in biological media. Complete: media supplemented with 10% foetal bovine serum (FBS); DEPC: diethylpyrocarbonate; DMEM: Dulbecco's Modified Eagle Medium; F12k: Ham's F-12K (Kaighn's) Medium; RPMI-1640: Roswell Park Memorial Institute 1640. Size distribution of BFO-NP after dispersion at 1 µg/mL in DEPC water solution (DEPC), DMEM medium (DMEM), DMEM medium supplemented with 10% FBS (complete DMEM), F12k medium (F12k), F12k medium supplemented with 10% FBS (complete F12k), RPMI 1640 medium (RPMI-1640) and RPMI 1640 medium supplemented with 10% FBS (complete RPMI-1640). All measurements were carried out three times at pH 7.4, then means and standard deviations (SD) were calculated.

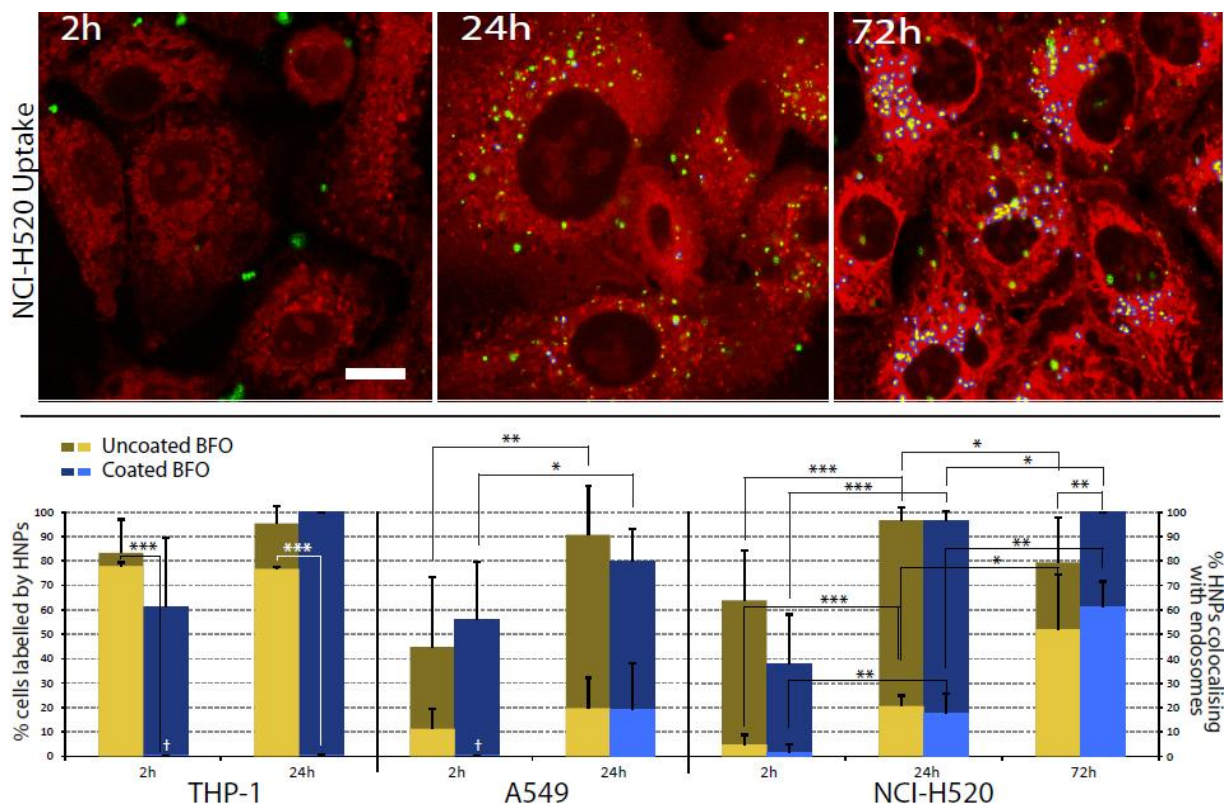

**Figure 3**: BFO-NP uptake in human-derived cell lines

First row: NCI-H520 cancer cells exposed to PEG-BFO-NP at 25 µg/mL for 2h, 24h and 72h. Green: SHG signal of BFO-NP; yellow: co-localization between the probe for lipid membranes and the particles; blue circles: highlight of HNP co-localizing with FM1-43FX. Scale bar: 10 µm.

Second row: % of labelled A549, NCI-H520 and THP-1 cells exposed to U-BFO-NP (brown bars) or PEG-BFO-NP (blue bars) at 25 µg/mL for 2h, 24h or 72h (only NCI-H520). Statistical comparisons were done using a Student's t-test: *p<0.05, **p<0.01, ***p<0.001. †: no HNP co-localizing with endosomes were observed.

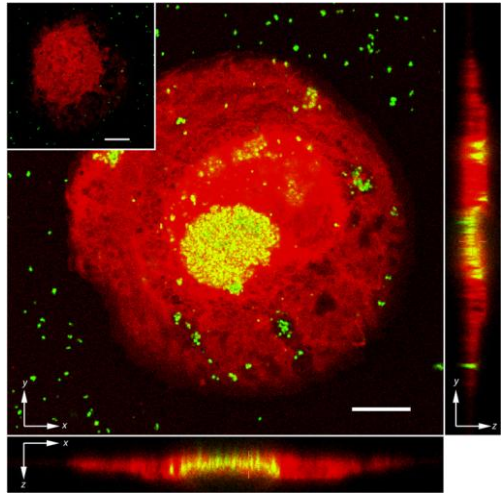

**Figure 4**: uptake of uncoated BFO-NP by activated THP-1 cells.

THP-1 cells were exposed for 24h at 25 µg/ml U-BFO-NP (green), then stained with a fluorescent probe for endosomes (red, FM1-43FX). The picture was extracted from a z-stack and shows 2 slice views centered at the yellow cross along its y z axis (right panel) and x z axis. Inset: Image of a THP-1 cell after 24h exposure to to 25 µg/ml of PEG-BFO-NP. Scale bars: 10 µm.

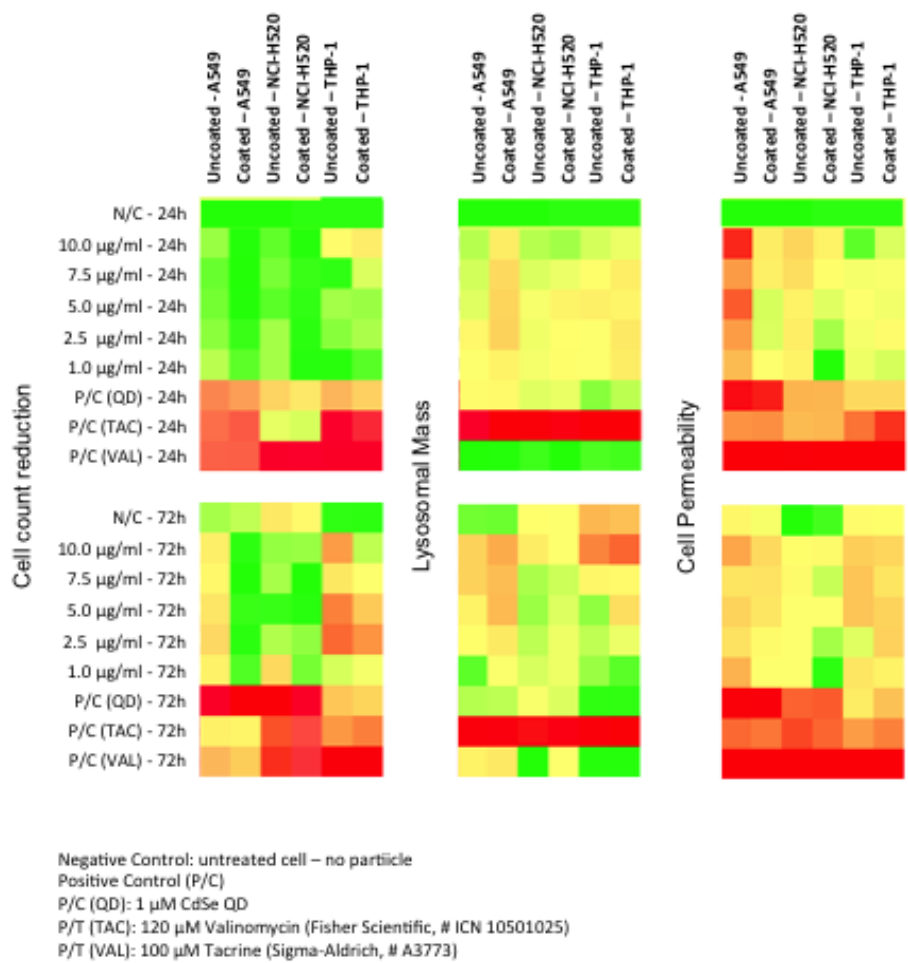

**Figure 5**: Heatmaps tables illustrating toxicity indicated parameters of BFO-NP in human-derived cell lines. Heatmaps were generated from the analysis of n= 3 experiments, each with triplicate wells for each of the parameters under investigation: cell count, lysosomal mass, cell permeability and nuclear area. Colorimetric gradient table spans from: Dark green: lower than 15% of maximum value measured; Bright green = 30%; Yellow = 50%; Bright orange: 60%; Dark orange = 75%; Red higher than 75% of maximum value. N/C untreated controls (negative) and P/C as positive controls such as QD: quantum dots, TAC: Tacrine, VAL: Valinomycin). Heatmap values are normalised using the per cent of the positive controls and, Z score was calculated as described in the statistical analysis section.

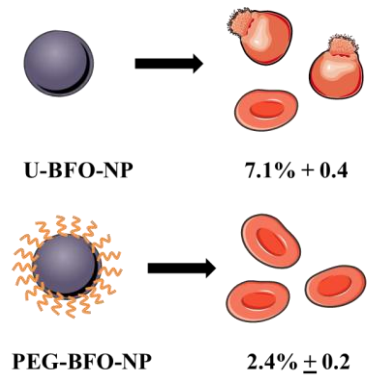

**U-BFO-NP**      7.1% + 0.4

**PEG-BFO-NP**      2.4% $\pm$ 0.2

**Figure 6**: Haemolytic effect of U-BFO-NP and PEG-BFO-NP.

HBRc were exposed to NP (25 µg/mL) for 2h. Results were expressed as % of haemolysis of NP-exposed HRBc compared to unexposed cells. Results are the means + SD of triplicates of two independent experiments. Values of U-BFO-NP were compared to those of PEG-BFO-NP by a Student's t-test (p<0.001). This picture was done using Servier Medical Art images.